\documentclass[final,1p,times]{elsarticle} 
\usepackage{graphicx}
\usepackage{amssymb} 
\usepackage{amsthm} 
\journal{Nuclear Physics A} 

\begin{document}

\begin{frontmatter} 

% Your Title - please insert
\title{Quantifying a Possibly Reduced
Jet-Medium Coupling of the sQGP at the LHC}

%% Single author (and collaboration) - please insert
\author{Barbara Betz$^{\rm a}$}
\address[auth1]{Institute for Theoretical Physics, Johann Wolfgang Goethe-University, 60438 Frankfurt am Main, Germany}
\author[auth2]{Miklos Gyulassy}
\address[auth2]{Department of Physics, Columbia University, New York, 10027, USA}

\begin{abstract} 
The nuclear modification factor $R_{AA}$ measured in Pb+Pb collisions
at the Large Hadron Collider (LHC) suggests that the jet-medium coupling in a 
Quark-Gluon Plasma at LHC energies is reduced as compared to energies reached at the
Relativistic Heavy Ion Collider (RHIC). We estimate the reduction factor
using a simple generic energy-loss model and find that the jet-medium coupling
at the LHC is approximately 10\% smaller than at RHIC. Moreover, we examine
different jet-energy loss prescriptions with $dE/dx\sim E^a$
and show that the measured momentum dependence of the nuclear modification
factor at the LHC rules out any model with $E^{a>1/3}$.
\end{abstract} 

\end{frontmatter} % do not change

%% linenumbers are useful for reviewing process
%\linenumbers

\section{Introduction}
The nuclear modification factor $R_{AA}$ measured at the Large
Hadron Collider (LHC) \cite{CMS,ALICE} suggests that
the jet-medium coupling is weaker than expected from fixed-coupling
extrapolations \cite{Horowitz} based on the data from
the Relativistic Heavy Ion Collider (RHIC) \cite{RHIC}.
While the rapid increase of the nuclear modification factor with
increasing transverse momentum $p_T$ can readily be understood
from generic perturbative physics \cite{CMS},
the reduced jet-medium coupling at the LHC tends to overquench 
those models constrained to RHIC data \cite{Horowitz}.

To quantify the reduction of the jet-medium coupling, we consider a 
generic jet-energy loss model \cite{us}
\begin{equation}
\hspace*{-3ex}
\;\;\;\frac{dP}{d\tau}(\vec{x}_\perp,\tau)= 
-\kappa[T(\vec{x}_\perp,\tau)] P^a(\tau) \, \tau^{z} \, T^ {c=2-a+z}(\vec{x}_\perp,\tau)\,,
\label{GenericEloss}
\end{equation}
where the energy loss per unit length or momentum loss per unit time
is proportional to a dimensionless coupling $\kappa$, the energy (momentum) dependence 
$P^a$, the path-length dependence $\tau^z$, and the local temperature dependence $T^c$.
The jet path $\vec{x}_\perp(\tau)=\vec{x}_0+\hat{n}(\phi)\tau$ is perpendicular to the 
beam axis from a production point $\vec{x}_0$ and moves in direction $\phi$
relative to the reaction plane. Here we assume the simplest boost-invariant
Bjorken hydrodynamics $T^3(\vec{x},\tau) = T^3(\vec{x},\tau_0)\tau_0/\tau$ down
to a freeze-out temperature of $T_f\sim 100$~MeV. The initialization time is
considered to be $\tau_0=1$~fm.

To compare different initial conditions, we consider Glauber participant
and binary initial geometries using a Monte Carlo model introduced
in Ref.\ \cite{us}. For CGC-like initial conditions however, we deform the
Glauber initial geometry via \cite{us}:
\begin{eqnarray}
x\rightarrow \sqrt{\frac{\langle x^2\rangle_{\rm CGC}}{\langle x^2\rangle_{\rm Gl}}} \;x, 
\quad y \rightarrow \sqrt{\frac{\langle y^2\rangle_{\rm CGC}}{\langle y^2\rangle_{\rm Gl}}}\; y\,.
\label{fkln1}
\end{eqnarray}

\begin{figure}[htbp]
\begin{center}
 \includegraphics[width=0.8\textwidth]{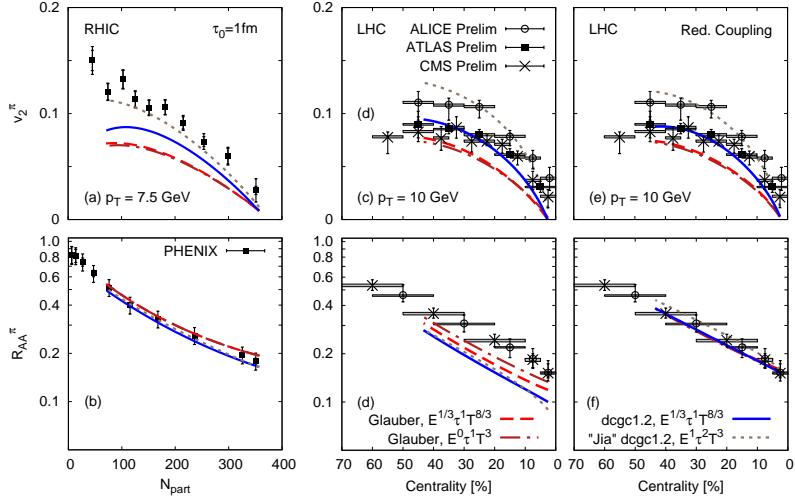}
\end{center}
\caption{The high-$p_T$ elliptic flow (upper panels) and the nuclear modification 
factor (lower panels) for RHIC (left) and LHC energies (middle and right panel),
considering different jet-energy loss prescriptions \cite{us} explained in the text. }
\label{fig1}
\end{figure}

Here, $\langle\circ\rangle$ denotes the geometric average at a given 
impact parameter $b$. The scaling factors are determined by fitting tabulated
Glauber and MC-KLN second moments. Assuming that the ratios of eccentricities ($\epsilon=e_2$) 
for Glauber and CGC initial conditions can be expressed via 
$\epsilon_{\rm CGC} = f\cdot\epsilon_{\rm Gl}$, we found that deformations with 
$f=1.2\pm 0.1$ reproduce the numerical MC-KLN tables very well \cite{us}. Therefore,
we refer to those CGC-like initial conditions as 'dcgc1.2'.

\section{Results and Discussions}

In the following, we are going to compare four different scenarios, see Eq.\ (\ref{GenericEloss}):
I. (a=0, z=1, Glauber), II. (a=1/3, z=1, Glauber), III. (a=1/3, z=1, dcgc1.2),
IV. (a=1, z=2, "Jia" dcgc1.2). 

The first three scenarios are based on a geometry with 14\% binary and 86\% participant collisions,
while the last scenario is based on pure binary collisions as in Ref.\ \cite{Jia}. Here, we
do not focus on the difference between the path-length dependencies as it was shown in a
most recent work \cite{Ficnarnew} that $z=1$ might be the correct
description for {\it both} a pQCD and an AdS/CFT energy loss prescription.

The first scenario $(a=0, z=1)$ coincides with the deep Landau-Pomeranchuk-Migdal
(LPM) pQCD limit, while the prescription with $(a=1/3, z=1)$ 
approximately describes both a pQCD and an AdS/CFT falling
string case \cite{ches1}: The $(a=1/3)$ power law is predicted to be the
lower bound of the power $a$ in a falling-string scenario. However,
an $(E/T)^{1/3}$ energy dependence is numerically similar to the 
logarithmic $\log(E/T)$ dependence predicted by fixed-coupling pQCD 
energy loss also relevant at LHC energies.

Fig.\ \ref{fig1} displays the comparison of the nuclear modification factor
(lower panel) and the high-$p_T$ elliptic flow (upper panel) as
a function of centrality for RHIC (left panel) and LHC energies (middle
and right panel) for the four scenarios discussed above. 

It can be seen from the lower left panel that once the coupling $\kappa$ is fixed
to reproduce the most central $R_{AA}$-data at RHIC energies, the
nuclear modification factor can be described well for all four models.
In contrast, the high-$p_T$ elliptic flow seems to be only well described
by the model with an energy loss of $dE/dx\sim E^1$.

Performing a straight extrapolation from RHIC to LHC energies (i.e.\ from the
left to the middle panel), a clear oversuppression of the nuclear modification
factor becomes visible. However, reducing the jet-medium coupling (right panel),
this oversuppression can be overcome. Please note that the reduced jet-medium
coupling does not have a big impact on the high-$p_T$ elliptic flow in the 
upper panel.

\begin{figure}[t]
\begin{center}
\begin{minipage}[t]{6.5cm}
 \includegraphics[width=1\textwidth]{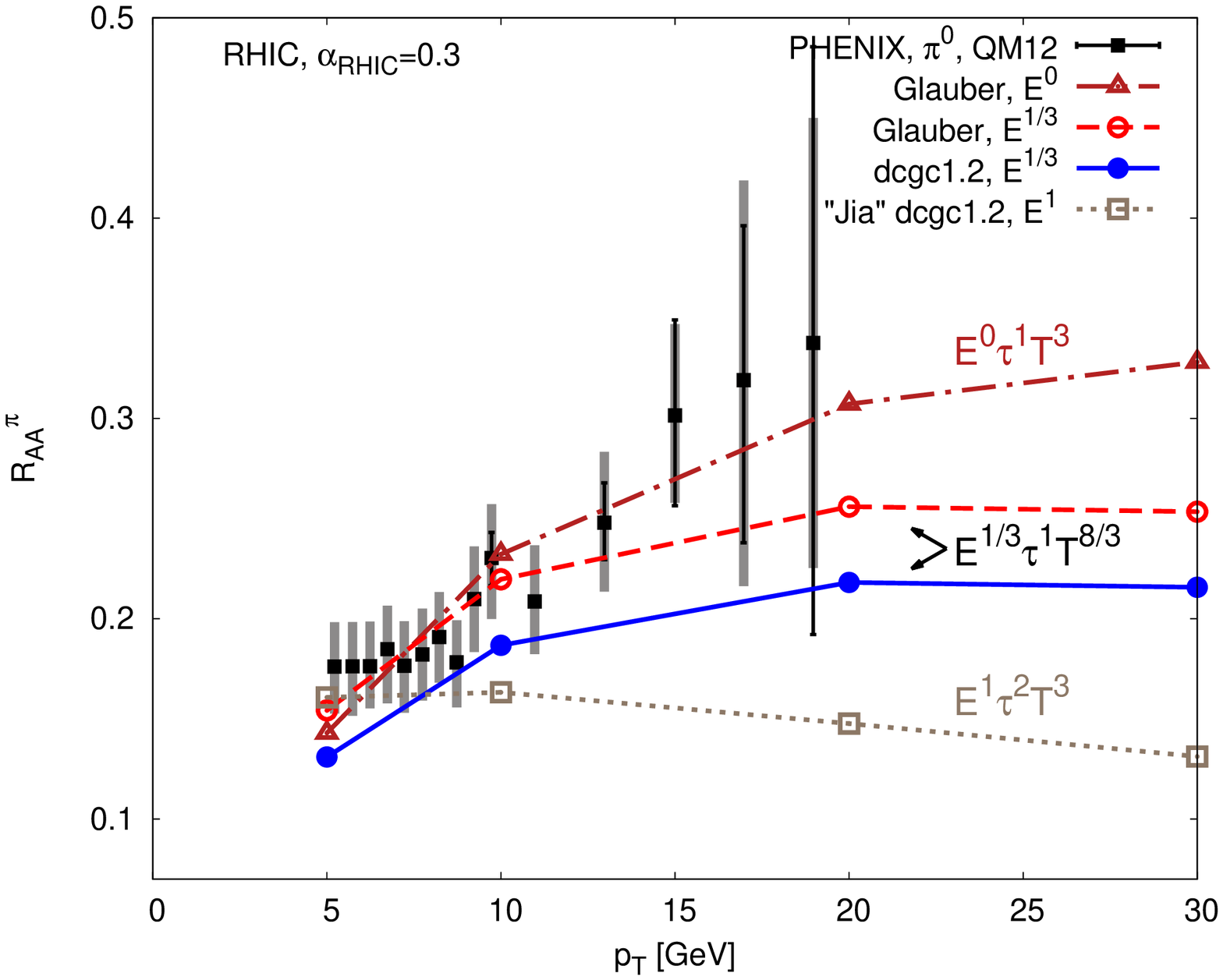}
\end{minipage}
\begin{minipage}[t]{6.5cm}
 \includegraphics[width=1\textwidth]{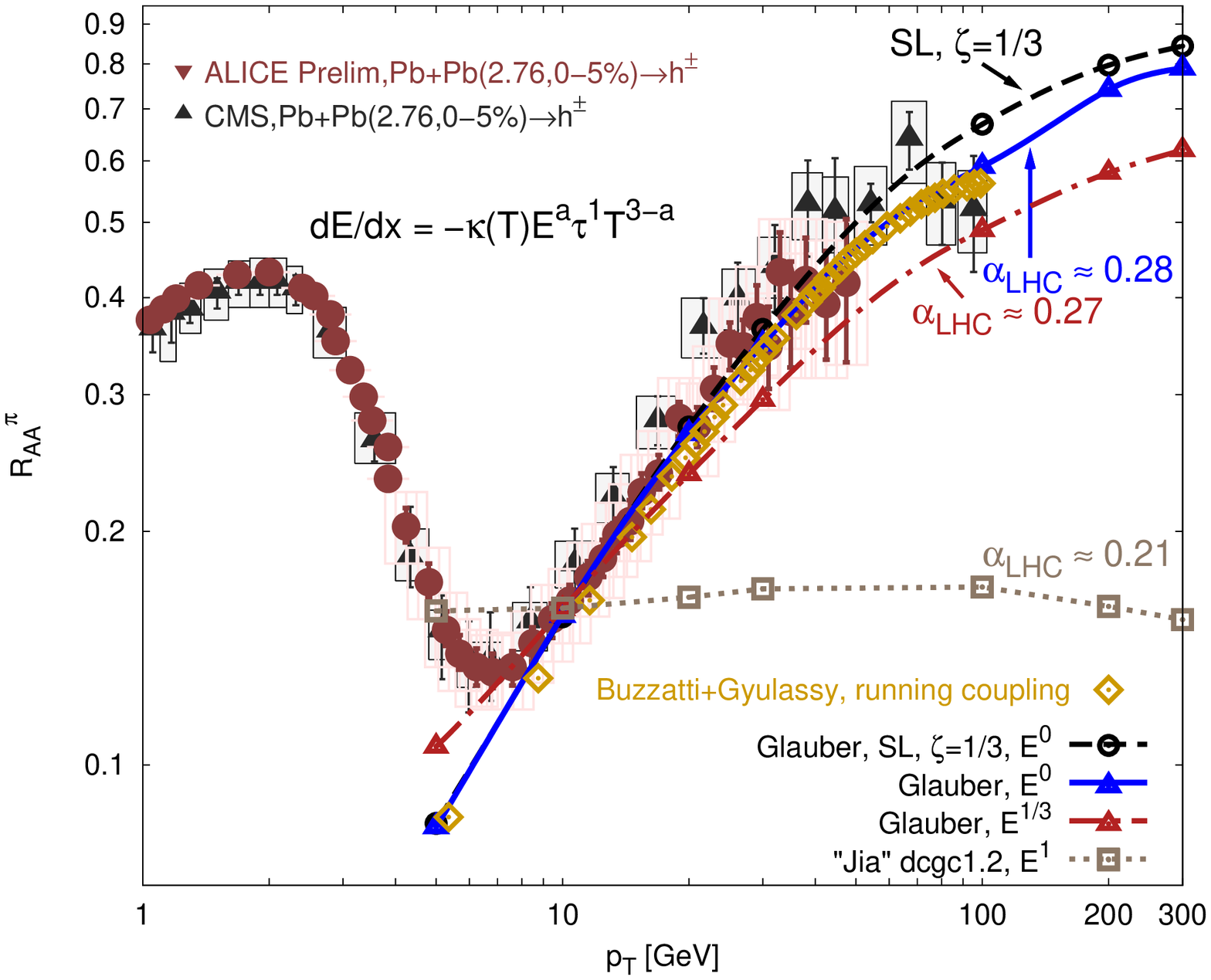}
\end{minipage}
\end{center}
\caption{The nuclear modification factor as a function of $p_T$ measured
at RHIC (left panel) \cite{RHIC} and at LHC energies (right panel) \cite{CMS,ALICE}.
The lines correspond to different jet-energy loss prescriptions \cite{us} explained in the text.}
\label{fig2}
\end{figure}

To quantify the magnitude of reduction, we recall that for a pQCD radiative
energy loss $\kappa \propto \alpha^3$. Thus, the strong coupling scales from 
RHIC to LHC as
\begin{eqnarray}
\alpha_{\rm LHC}&=& (\kappa_{\rm LHC}/\kappa_{\rm RHIC})^{1/3} \; \alpha_{\rm RHIC}\,,
\end{eqnarray}
where $\alpha_{\rm RHIC}\sim 0.3$. Inserting all values used for the couplings 
$\kappa_{\rm RHIC}$ and $\kappa_{\rm LHC}$ \cite{us}, we obtain a moderate reduction
of the running coupling of $\kappa_{\rm LHC}\sim0.24 - 0.28$. Note 
that due to the relation $\kappa \propto \alpha^3$, a jet-medium
reduction at the LHC is directly coupled to a reduction in the coupling constant
$\kappa$.

Considering hwoever the effective jet-medium coupling in a falling-string
scenario, $\kappa\propto \sqrt{\lambda}$ is related to the 
square root of the t'Hooft coupling $\lambda=g_{YM}^2 N_c$. In this case,
\begin{eqnarray}
\lambda_{\rm LHC}&=& (\kappa_{\rm LHC}/\kappa_{\rm RHIC})^2 \; \lambda_{\rm RHIC}
\end{eqnarray}
yields to a reduction of the t'Hooft coupling $\kappa_{\rm LHC}$ up to a 
factor of four ($\kappa_{\rm RHIC}=20$) \cite{us}.
It is not yet clear if the current models allow for such a strong breaking of the
conformal symmetry.

Fig.\ \ref{fig2} compares the nuclear modification factor as a function of
$p_T$ for RHIC (left) and LHC (right) energies. Both Figures show that the model 
with $dE/dx\sim E^1$ that best fit both the $R_{AA}$ and the high-$p_T$ elliptic
flow at RHIC (see left panel of Fig. \ref{fig1}) does not follow the trend 
of the data at the LHC.

However, for those models with $dE/dx\sim E^0$ and $dE/dx\sim E^{1/3}$, there
is a clear discrepancy between the measured data for the elliptic flow $v_2$ and
the data at RHIC (see again upper left panel of Fig. \ref{fig1}). 

The reason is that the RHIC data are considered for an intermediate $p_T=7.5$~GeV.
As demonstrated in Ref.\ \cite{Molnar}, parton coalescence might enhance the
measured elliptic flow in such an intermediate $p_T$-range by a factor of $2-3$. Thus,
one should {\it not} expect that pure jet fragmentation and absorption models
like the one discussed here fully describes the intermediate $p_T$-range.

On the other hand, Fig.\ \ref{fig2} clearly shows that the prescriptions with
$dE/dx\sim E^0$ and $dE/dx\sim E^{1/3}$ describe the trend of the nuclear modification
factor as a function of $p_T$.

Additionally to the scenarios mentioned earlier, the right panel of Fig.\ \ref{fig2} 
shows the results based on (1) a temperture-dependent coupling as introduced in Ref.\
\cite{LS} by Shuryak and Liao (SL) (dashed black line) and (2) on
the CUJET code by Buzzatti and Gyulassy \cite{Buzzatti} (golden diamonds) which is based on the 
Gyulasssy-Levai-Vitev (GLV) approach and includes a running coupling. 
In the first case (SL model), it is assumed that the coupling is enhanced 
in a mixed phase ($\kappa_2$, $113<T<173$~MeV) as compared to the QGP phase 
($\kappa_1$, $T>173$~MeV) by a factor of $\zeta=\kappa_1/\kappa_2=1/3$. 

As the figure shows, all models considering either $dE/dx\sim E^0$ or $dE/dx\sim E^{1/3}$ 
follow the trend of the measured data up to $p_T\sim 20-50$~GeV. However, 
none of the models discussed here leads to a plateau-like structure of
the nuclear modification factor for $p_T>100$~GeV as indicated by the data \cite{CMS,ALICE,CMS2}.

\section{Summary}
Using a simple generic energy-loss model, we estimated the reduction factor
of the jet-medium coupling between RHIC and LHC and found a 10\% moderate reduction
of the running coupling. Moreover, we showed that all models with $dE/dx\sim E^{a<1/3}$
follow the trend of the measured nuclear modification factor {\it both}
at RHIC and LHC energies.

\section{Acknowledgments}

B.B.\ is supported by the Alexander von Humboldt foundation. M.G.\ and
B.B.\ acknowledge support from DOE under Grant No.\ 
DE-FG02-93ER40764. The authors thank A. Buzzatti and G.\ Torrieri 
for fruitful and stimulating discussions.

%\section*{References}


\begin{thebibliography}{00} 
\bibitem{CMS} S.~Chatrchyan {\it et al.}  [CMS Collaboration], Eur.\ Phys.\ J.\ C {\bf 72}, 1945 (2012).
\bibitem{ALICE} B.~Abelev {\it et al.}  [ALICE Collaboration], arXiv:1208.2711 [hep-ex].
\bibitem{Horowitz} W.~A.~Horowitz and M.~Gyulassy, Nucl.\ Phys.\ A {\bf 872}, 265 (2011).
\bibitem{RHIC} A.~Adare {\it et al.}  [PHENIX Collaboration], arXiv:1208.2254 [nucl-ex].
\bibitem{us} B.~Betz, M.~Gyulassy and G.~Torrieri, Phys.\ Rev.\ C {\bf 84}, 024913 (2011); 
             B.~Betz and M.~Gyulassy, Phys.\ Rev.\ C {\bf 86}, 024903 (2012).
\bibitem{Jia} A.~Drees, H.~Feng, and J.~Jia, Phys.\ Rev.\  C {\bf 71}, 034909 (2005);
              J.~Jia and R.~Wei, Phys.\ Rev.\  C {\bf 82}, 024902 (2010).
\bibitem{Ficnarnew} A.~Ficnar, arXiv:1201.1780.
\bibitem{ches1} S.~S.~Gubser, D.~R.~Gulotta, S.~S.~Pufu and F.~D.~Rocha, J.~High Energy Phys.\ {\bf 10}, 052 (2008).
\bibitem{Molnar} D.~Molnar, J.\ Phys.\ G {\bf 30}, S235 (2004).
\bibitem{LS} J.~Liao and E.~Shuryak, Phys.\ Rev.\ Lett.\  {\bf 102}, 202302 (2009).
\bibitem{Buzzatti} A.~Buzzatti and M.~Gyulassy, Phys.\ Rev.\ Lett.\  {\bf 108}, 022301 (2012).
\bibitem{CMS2} CMS Collaboration, CMS-PAS-HIN-12-004.

\end{thebibliography}
\end{document}